\documentclass[fleqn,usenatbib]{mnras}

\usepackage{newtxtext,newtxmath}

\usepackage[T1]{fontenc}

\DeclareRobustCommand{\VAN}[3]{#2}
\let\VANthebibliography\thebibliography
\def\thebibliography{\DeclareRobustCommand{\VAN}[3]{##3}\VANthebibliography}

\usepackage{graphicx}	
\usepackage{amsmath}	

\usepackage{longtable}
\usepackage{supertabular,booktabs}
 \usepackage{cancel}
\usepackage{tikz}
\usetikzlibrary{matrix}
\usepackage{bm}

\newcommand{\tim}{\textcolor{black}}

\makeatletter
\newcommand\footnoteref[1]{\protected@xdef\@thefnmark{\ref{#1}}\@footnotemark}
\makeatother

\title[Reduction of supernova LCs by vector GPs]{Reduction of supernova light curves by vector Gaussian processes}

\author[M. V. Kornilov et al.]{
Matwey V. Kornilov$^{1,2}$,
T. A. Semenikhin$^{1,3}$\thanks{E-mail: ofmafowo@gmail.com},
M. V. Pruzhinskaya$^{1,4}$
\\
$^{1}$Lomonosov Moscow State University, Sternberg astronomical institute, Universitetsky pr. 13, Moscow 119234, Russia;\\
$^{2}$National Research University Higher School of Economics, 21/4 Staraya Basmannaya Ulitsa, Moscow, 105066, Russia; \\
$^{3}$Faculty of Space Research, Lomonosov Moscow State University, Leninsky Gori 1 bld. 52, Moscow 119234, Russia; \\
$^{4}$Universit\'e Clermont Auvergne, CNRS/IN2P3, LPC, F-63000 Clermont-Ferrand, France
}

\date{Accepted XXX. Received YYY; in original form ZZZ}

\pubyear{2023}

\begin{document}
\label{firstpage}
\pagerange{\pageref{firstpage}--\pageref{lastpage}}
\maketitle

\begin{abstract}
Bolometric light curves play an important role in understanding the underlying physics of various astrophysical phenomena, as they allow for a comprehensive modeling of the event and enable comparison between different objects. However, constructing these curves often requires the approximation and extrapolation from multicolor photometric observations. In this study, we introduce vector Gaussian processes as a new method for reduction of supernova light curves. This method enables us to approximate vector functions, even with inhomogeneous time-series data, while considering the correlation between light curves in different passbands. We applied this methodology to a sample of 29 superluminous supernovae (SLSNe) assembled using the Open Supernova Catalog. Their multicolor light curves were approximated using vector Gaussian processes. Subsequently, under the black-body assumption for the SLSN spectra at each moment of time, we reconstructed the bolometric light curves. The vector Gaussian processes developed in this work are accessible via the Python library {\tt gp-multistate-kernel} on GitHub. Our approach provides an efficient tool for analyzing light curve data, opening new possibilities for astrophysical research.
\end{abstract}




\section{Introduction}

The astrophysical processes have different time duration --- from a few seconds (e.g., gamma-ray bursts) to billions of years (e.g., galaxy mergers). The events with a duration not exceeding the length of a human's life are of particular interest since they can be thoroughly scrutinized by the community from the beginning to the very end. However, even if we perform continuous observations of an astronomical object, the obtained data are often unevenly distributed in time, since the success of measurements highly depends on factors that are beyond our control, such as weather conditions. Many fundamental and applied tasks require a time-homogeneous input data set. A commonly used technique to obtain such time series is to approximate them with Gaussian processes (GPs,~\citealt{3569}).

GP is a particularly efficient method of placing a prior distribution over the space of functions which is widely used in astrophysics for regression and classification problems. In~\citet{2019AJ....158..257B} GPs have been used to predict smooth models for sparsely sampled multicolor light curves (LCs) of the PLAsTiCC data set \citep{2020arXiv201212392H}. These models were then further used for classification. \citet{2017ApJ...840...49C} demonstrated the successful application of Gaussian processes to model spectra of single-lined and double-lined spectroscopic binaries. \citet{10.1111/j.1365-2966.2011.19915.x} introduced GPs for the modelling and removal of systematics in a planetary transit observation.
GPs can be useful for modelling stochastic variability in active galactic nuclei optical and radio emission~\citep{1992ApJ...385..404P}. In~\citet{refId0} GPs were applied to approximate SN Ia LCs on a fixed time grid. Also, the use of principal component analysis together with GPs allows to identify galaxies overlapping along the same line of sight~\citep{Buchanan_2022} and to solve the star-galaxy image separation task~\citep{Muyskens_2022}.

Many open-source tools have been developed to implement GPs in astrophysics (see section 6 of~\citealt{rev} for a review). Among the most popular, there are the Python libraries {\tt scikit-learn}\footnote{\url{https://scikit-learn.org/stable/modules/gaussian_process.html}} and {\tt george}\footnote{\url{https://george.readthedocs.io/}} \citep{https://doi.org/10.48550/arxiv.1403.6015}. However, for some tasks the researchers have to develop their custom GPs. 

Traditionally, machine learning algorithms require data of fixed dimensionality. Therefore, solving the anomaly detection problem in photometric time-series, a method to restore the missing parts of LCs in one band based on the observations in another is needed. To do so, \citet{2019MNRAS.489.3591P} used the so-called vector Gaussian processes\footnote{\label{vgp}\url{https://gp.snad.space/}}. This method allows us to approximate the vector functions even if the training data are inhomogeneous in time, and therefore to take into account the correlation between LCs in different passbands. In this work, we define the vector GPs and apply this method to reconstruct the bolometric\footnote{Throughout this paper, despite using the term "bolometric light curve", we are in fact referring to quasi-bolometric light curves, as they are reconstructed from a limited range of wavelengths.} light curves of superluminous supernova (SLSN) sample from the Open Supernova Catalog\footnote{\label{osc}\url{https://sne.space/}} (OSC, ~\citealt{2017ApJ...835...64G}).

Being significantly brighter ($L_{\rm max} \gtrsim 10^{44}$ erg~s$^{-1}$) than normal supernovae, SLSNe demonstrate a high diversity in their photometrical and spectroscopical behavior \citep{Gal_Yam_2012,Gal_Yam_2019,Moriya_2018}. If there are no hydrogen lines in the spectrum, and the light curve rapidly evolves, an SLSN is assigned to type I (SLSN I); however, if it has a broad, slowly decreasing light curve, which could be explained by the radioactive decays, it belongs to SLSN~R ("R" for radioactive). Type II is used for hydrogen-rich events (SLSN~II) and shows the LCs of many different shapes. There are three main sources of energy that can explain the variety of observational properties and the high luminosity of SLSNe: the radioactive decay of a large amount of $^{56}\rm{Ni}$ ($>$ 5~$M_\odot$; e.g., \citealt{2015MNRAS.452.3869N}); energy input from a central engine --- magnetar~\citep{2010ApJ...719L.204W,Kasen_2010} or accretion onto a newly formed black hole~\citep{Dexter_2013}; interaction of ejecta with dense circumstellar medium associated with pre-explosion mass loss of the progenitor (e.g.,~\citealt{Chevalier_2011}). However, none of these hypotheses can fully explain the shape of LCs and spectra of superluminous supernovae. Having a unified sample of SLSN LCs will allow us to compare objects one by one, as well as to study the population as a whole, which can shed light on power input mechanism of SLSNe.

The rest of the paper is structured as follows: in Section~\ref{sample}, we describe the data presented in the OSC and selection criteria to SLSN sample. Section~\ref{methods} is devoted to mathematical description of the vector Gaussian processes (\ref{gp}) and their implementation for multicolor light-curve approximation (\ref{ap_gp}). Section~\ref{bol} presents a method for constructing bolometric light curves from approximated multicolor light curves. We compare our results with other works in Section~\ref{disc} and briefly conclude in Section~\ref{concl}.

\section{Sample of superluminous supernovae}
\label{sample}

\begin{figure}
        \center{\includegraphics[width=\columnwidth]{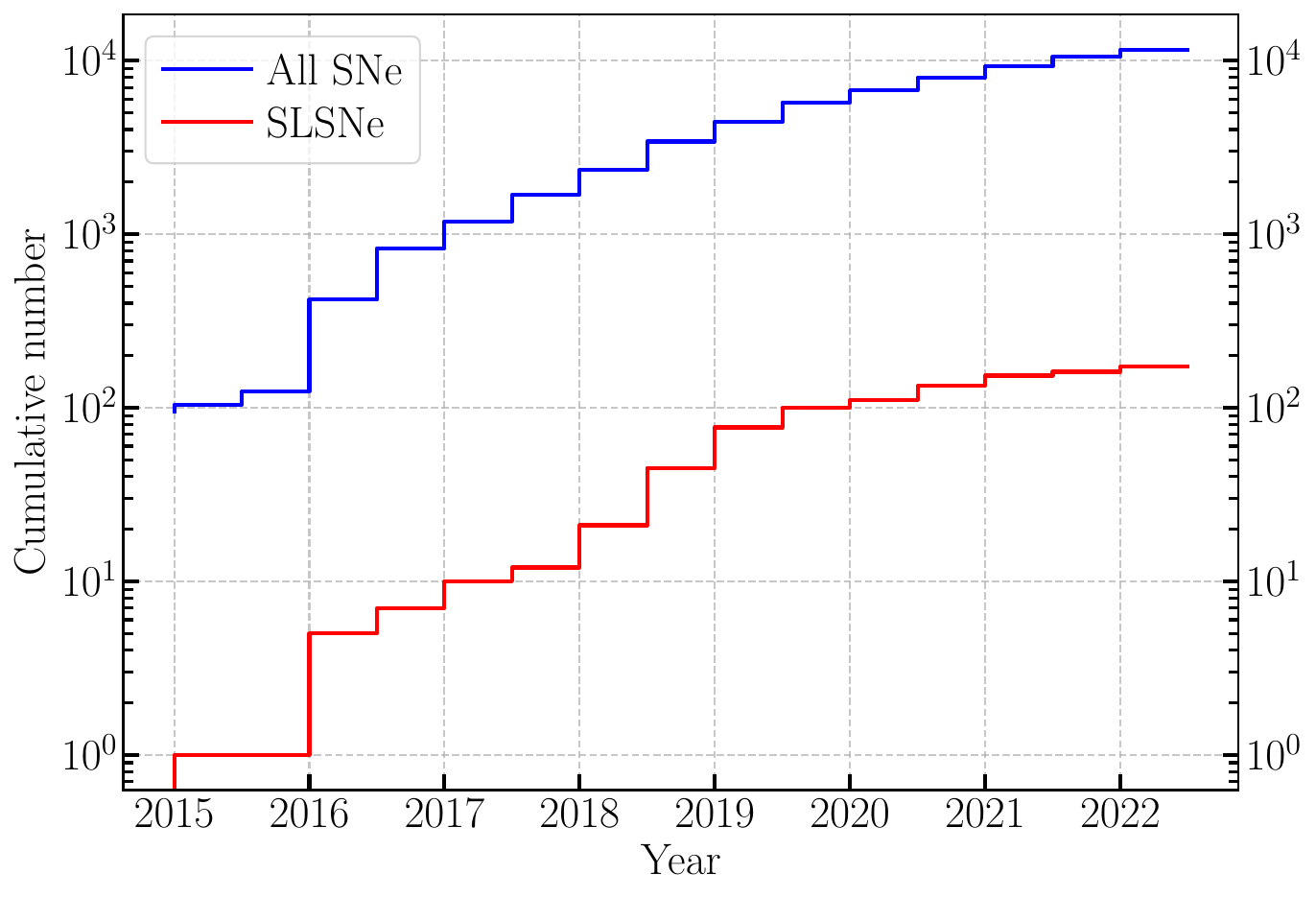}}
        \caption{Cumulative distribution of spectroscopically-confirmed supernovae by years according to TNS. Red line~--- SLSNe, blue line~--- all supernovae.}
        \label{fig:intr:1} 
\end{figure}

At the moment, over 100 SLSNe have been discovered and classified~\citep{Nicholl_2021,2023ApJ...943...41C}. Fig.~\ref{fig:intr:1} shows the cumulative distribution of spectroscopically-confirmed supernovae over the years according to Transient Name Server\footnote{\url{https://www.wis-tns.org/}} (TNS): about ten thousand SNe of all classes and one hundred of SLSNe. It can be seen that the number of discovered SLSNe has greatly increased after 2016. This is explained by the fact that earlier surveys primarily focused on observing massive galaxies, whereas it has been found that SLSNe typically occur in dwarf galaxies~\citep{Nicholl_2021,Gal_Yam_2019}.

To make a sample of SLSNe, we extracted the data from the Open Supernova Catalog\footnote{The catalogue is not supported anymore. The last update is from April 8, 2022. The archival data can be found on GitHub \href{https://github.com/astrocatalogs/supernovae}{here}.}~\citep{2017ApJ...835...64G} using the {\tt snad}\footnote{\url{https://github.com/snad-space/snad}} package. The catalog represents a compilation of different data sources (e.g., Pan-STARRS~\citep{2010SPIE.7733E..0EK,2016arXiv161205560C}, SDSS Supernova Survey~\citep{Sako2018},  Sternberg Astronomical Institute Supernova Light Curve Catalogue~\citep{2005yCat.2256....0T}, Supernova Legacy Survey ~\citep{2005ASPC..339...60P,2006A&A...447...31A}, MASTER Global Robotic Net~\citep{2010AdAst2010E..30L}, All-Sky Automated Survey for Supernovae \citep{2019MNRAS.484.1899H}), including the individual publications. The data presented in the catalog are inhomogeneous in terms of photometric systems, number of observations, number of passbands, observational cadence, and classification reliability. In total, the OSC contains 239 objects classified as superluminous supernovae.

To ensure that our final sample consists of SLSNe only, we selected those objects that have at least one spectrum in the catalog, which amounted to 71 out of 239. In order to reconstruct the bolometric light curves with better accuracy, we did not limit ourselves to a single magnitude system: we extracted all photometric data available in both the $UBVRI$ Vega system and the $ugriz$/$u'g'r'i'z'$ AB system. For the further analysis, we assume that the transmissions of the $ugriz$ filters are sufficiently close to those of the $u'g'r'i'z'$ filters~\citep{1996AJ....111.1748F,2006AN....327..821T,2007ASPC..364...91S}. The only requirement we impose is to have observations in at least three different passbands among these systems (see Table~\ref{tab}). In total, we found 45 SLSNe satisfying these criteria. Then, for each of these SLSNe, we identified the date corresponding to the brightest magnitude among all passbands $MJD_{\rm max}$, and discarded all photometric observations that were not within the range $[MJD_{\rm max} - 50; MJD_{\rm max} + 150]$.

\section{Methods}
\label{methods}

\subsection{Overview}
\label{overview}
Traditionally, Gaussian processes are used to fit the light curves in each passband independently. In such case, the model does not take into account the correlation between different passbands, and the LCs with few of observations and/or absence of photometric points before/after the maximum light are not fitted satisfactorily.  However, for some tasks, e.g. in order to construct the bolometric LCs, the maximum photometric information is preferred for use. 

There are different methods that allow us to take into account the correlation between passbands. \citet{2019AJ....158..257B} presents a set of techniques for photometric classification of astronomical transients and variables trained and evaluated on the Photometric LSST Astronomical Time-Series Classification Challenge (PLAsTiCC, \citealt{Kessler_2019}). Gaussian process regression were used to model arbitrary LCs in all passbands simultaneously, thus to “augment” the training set by generating new versions of the original LCs covering a range of redshifts and observing conditions. GPs were applied as a function of two variables --- approximated in both time and in wavelength simultaneously --- which made it possible to take into account the correlation between passbands. \citet{2019AJ....158..257B}  found the central wavelength for each filter and modeled light curves in this filter by fitting to central wavelength, which assumes the object spectrum is convolved with a broad filter. A similar approach to Gaussian processes, among other approximation methods, has been considered by \citet{2022arXiv220907542D}, but instead of using the wavelength, they employed its logarithm.

In this work, we propose a method to approximate the vector-valued functions, i.e. mappings from $\mathbb{R}$ to $\mathbb{R}^n$. The approximation is performed in terms of time and a set of passbands; no additional assumption about the shape of filter transmission curve is required. Below we describe the mathematical background of the method and its implementation to multicolor LCs of SLSNe.

\subsection{Vector Gaussian process}
\label{gp}

When GPs are employed in approximation, usually only a single approximated quantity is considered. Formally, we consider a mapping $\mathbb{R}^{m} \rightarrow \mathbb{R}$ constructed from some stochastic framework.
In our case, the original set~(time $t$) is one-dimensional itself (thus, $m=1$), so we can write the following equation for $y$ being approximated as a function of $t$:

\begin{equation}
\label{eq:gp:1}
y\left(t\right) = {\mathrm M}\left[\nu(t)\right] \equiv \int \nu p_{\nu}(\nu, t; \bm{\theta}) d \nu,
\end{equation}
where $\nu(t)$ is a random variable with mean $y(t)$, ${\mathrm M}\left[\cdot\right]$ denotes the mean operator, $p_{\nu}$ is a probability density function for $\nu$ with parameters $t$ and $\bm{\theta}$. Particular Gaussian processes parameters, $\bm{\theta}$, are of the mathematical expectation function $\mu(t)$ and the kernel $K(t_1, t_2)$ in GP approximation frameworks.
The probability density function has the following natural form:
\begin{equation}
\label{eq:gp:2}
p_{\nu}(\nu, t; \bm{\theta}) = p_{\nu}(\nu, t| \hat\nu_1, t_1, \dots, \hat\nu_N, t_N;\bm{\theta}(\hat\nu_1, t_1, \dots, \hat\nu_N, t_N)),
\end{equation}
where $p_{\nu}(\cdot | \cdot)$ denotes a conditional probability density function, $\hat\nu_i$ is the value measured at time $t_i$, parameters $\bm{\theta}$ are obtained using the maximum likelihood principle.

If we want to approximate multivariate vector $\bm{y}(t)$ then the traditional approach will work: instead of approximating multivariate vectors we simply consider a number of independent approximation problems for each vector component.
However, the naive approach may be generalised further.
It is obvious that the probability density function for an arbitrary random vector is not a multiplication of separate probability density functions for each individual component in the general case. Thus, instead of equation~\eqref{eq:gp:1} we have the following:
\begin{equation}
\label{eq:gp:3}
\bm{y}\left(t\right) = {\mathrm M}\left[\bm{\upnu}(t)\right] \equiv \int \bm{\upnu} p_{\bm{\upnu}}(\bm{\upnu}, t; \bm{\theta}) d \bm{\upnu},
\end{equation}
where boldfaced $\bm{\upnu}$ denotes multivariate vector and $d \bm{\upnu}$ stands for multidimensional integration.
Multivariate probability density function $p_{\bm{\upnu}}$ takes into account correlation between the components of~$\bm{\upnu}$.

Without loss of generality, let us consider a two-point probability density function $p_{\bm{\upnu}}(\bm{\upnu_1}, t_1, \bm{\upnu_2}, t_2; \bm{\theta})$:
\begin{equation}
\label{eq:gp:4}
\begin{split}
p_{\bm{\upnu}}(\bm{\upnu_1}, t_1, \bm{\upnu_2}&, t_2; \bm{\theta}) = \frac{1}{(2 \pi)^{\rm D/2} (\det \left|\Sigma\right|)^{1/2}} \\
&
\cdot \exp\left(-\frac{1}{2} \left(
		\left[\begin{aligned} \bm{\upnu_1} \\ \bm{\upnu_2} \end{aligned}\right] -
		\left[\begin{aligned} \bm{\mu_1} \\ \bm{\mu_2} \end{aligned}\right]
	,
	\Sigma^{-1}\left(
		\left[\begin{aligned} \bm{\upnu_1} \\ \bm{\upnu_2} \end{aligned}\right] -
		\left[\begin{aligned} \bm{\mu_1} \\ \bm{\mu_2} \end{aligned}\right]
	\right)
\right)\right),
\end{split}
\end{equation}
where $\bm{\mu_i} = \bm{\mu}(t_i)$ is the mean, determined by the expectation function (if the form of the function being approximated is unknown {\it a priori}, it is assumed to be equal to zero); $\Sigma$ is the covariance matrix; ${\rm D}$ is the vector $\left[\begin{aligned} \bm{\upnu_1} \\ \bm{\upnu_2} \end{aligned}\right]$ dimension; $\left[\cdot\right]$ denotes vector concatenation; $\left(\cdot,\cdot\right)$ denotes inner product.

The covariance matrix $\Sigma$ has the following special block form:
\begin{equation}
\label{eq:gp:5}
\Sigma = \left(\begin{aligned}
\Sigma_{\rm d} & \Sigma_{\rm s} \\
\Sigma_{\rm s}^{\rm T} & \Sigma_{\rm d} \\
\end{aligned}\right),
\end{equation}
where $\Sigma_{\rm d}$ is the covariance matrix for vector components and $\Sigma_{\rm s}$ is the convariance block for time-based covariance.
The former is demonstrated in Appendix~\ref{ap:1}, we integrate over $\bm{\upnu_2}$ which is equivalent to dropping out all matrix blocks except for the top left block $\Sigma_d$.
The result is the convariance matrix for the single-point probability density function of $\bm{\upnu_1}$. In Appendix~\ref{ap:2} the case with mutually independent components is considered.

In order to derive consistent form for $\Sigma$ in the general case, let us recall the following well-known result. Let $\bm{\epsilon}$ be a random vector with zero mean and unit covariance matrix $I$ and let ${\mathrm R}$ be a lower-triangular square matrix, then it is easy to demonstrate that ${\mathrm R} \bm{\epsilon}$ is a random vector with zero mean and covariance matrix ${\mathrm R} {\mathrm R}^T$.

Let us represent $\bm{\upnu} = {\mathrm R} \bm{\epsilon}$, where $\bm{\epsilon}$ has pair-wise uncorrelated components. It follows that
\begin{equation}
\label{eq:gp:7}
\Sigma_{s,(ij)} = \sum_{k=0}^{k \le \min\{i,j\}} R_{ik} R_{jk} K_k(t_1,t_2),
\end{equation}
where $K_k(t_1, t_2)$ is a covariance for $\epsilon_k$ at time moments $t_1$ and $t_2$, referred to as the ``kernel'' in Gaussian processes approximation framework.

Now we are able to construct the whole covariance matrix $\Sigma$ if we chose kernels $K_k$ for generating process $\bm{\epsilon}$.
Then, $\Sigma$ should be used in conjunction with a maximum likelihood technique to obtain unknown modeling parameters, such as elements of matrix ${\mathrm R}$ and kernel parameters.

Note that until now we considered the case when all components of $\bm{y}$ are known at the every time moment $t$.
However, this assumption may be relaxed further.
As demonstrated in Appendix~\ref{ap:1}, we can drop out the columns and rows corresponding to unknown $\bm{y}$ components. For example, let the vectors $\bm{\upnu_1}$ and $\bm{\upnu_2}$ (from equation~\eqref{eq:gp:4}) have two dimensions~(for instance, encoding magnitudes in two different passbands), such that $\bm{\upnu_1} = \left[\nu_1, \nu_2\right]^{\rm T}$ and $\bm{\upnu_2} = \left[\nu_3, \nu_4\right]^{\rm T}$. Thus, the mean vector and the covariance matrix will be:

\begin{equation}
\label{eq:gp:8}
\bm{\mu} = \left[ \mu_1, \mu_2, \mu_3, \mu_4 \right]^{\rm T},
\end{equation}

\begin{equation}
\label{eq:gp:9}
\Sigma = \left(\begin{array}{cccc}
\Sigma_{11} & \Sigma_{12} & \Sigma_{13} & \Sigma_{14} \\
\Sigma_{21} & \Sigma_{22} & \Sigma_{23} & \Sigma_{24} \\
\Sigma_{31} & \Sigma_{32} & \Sigma_{33} & \Sigma_{34} \\
\Sigma_{41} & \Sigma_{42} & \Sigma_{43} & \Sigma_{44} \\
\end{array}\right).
\end{equation}
Now suppose that we have no observations of the first component of the vector $\bm{\upnu_1}$ and of the second component of the vector $\bm{\upnu_2}$. Then, according to Appendix~~\ref{ap:1}, the mean vector and the covariance matrix (equations~\eqref{eq:gp:8},~\eqref{eq:gp:9}) will take the form:
\begin{equation}
\label{eq:gp:10}
\bm{\mu} = \left[ \cancel{\mu_1}, \mu_2, \mu_3, \cancel{\mu_4} \right]^{\rm T} = \left[ \mu_2, \mu_3 \right]^{\rm T},
\end{equation}

\begin{equation}
\label{eq:gp:11}
\Sigma = \begin{tikzpicture}[baseline]
     \matrix (magic) [%
      matrix of math nodes,
      text width=5mm,
      left delimiter=(,
      right delimiter=)
    ]{%
\Sigma_{11} & \Sigma_{12} & \Sigma_{13} & \Sigma_{14} \\
\Sigma_{21} & \Sigma_{22} & \Sigma_{23} & \Sigma_{24} \\
\Sigma_{31} & \Sigma_{32} & \Sigma_{33} & \Sigma_{34} \\
\Sigma_{41} & \Sigma_{42} & \Sigma_{43} & \Sigma_{44} \\
};
\draw[thick,red] (magic-1-1.north) -- (magic-4-1.south);
    \draw[thick,red] (magic-1-1.west) -- (magic-1-4.east);
    \draw[thick,blue] (magic-1-4.north) -- (magic-4-4.south);
    \draw[thick,blue] (magic-4-1.west) -- (magic-4-4.east);
  \end{tikzpicture} = %
  \left(\begin{array}{cc}
\Sigma_{22} & \Sigma_{23} \\
\Sigma_{32} & \Sigma_{33}  \\
\end{array}\right).
\end{equation}

Now that we have defined the covariance matrix $\Sigma$ in a general form, we can write an expression for the mathematical expectation under the condition of observations. Let vector $\bm{\upnu_2}$ (in terms of equation~\eqref{eq:gp:4}) be our observations at the time $t_2$ and we want to obtain an approximation of vector $\bm{\upnu_1}$. Then it can be shown~(see~\cite{VONMISES1964368}, section 9.3) that the conditional distribution of random variable $\bm{\upnu_1}$ takes the following form:
\begin{equation}
    \label{eq:gp:12}
    \bm{\upnu_1} | \bm{\upnu_2} \sim N\left( \bm{\mu_1} + \Sigma_{\rm s} \Sigma_{\rm d}^{-1}( \bm{\upnu_2} - \bm{\mu_2}), \; \Sigma_{\rm d} - \Sigma_{\rm s} \Sigma_{\rm d}^{-1} \Sigma_{\rm s}^{\rm T} \right).
\end{equation}
Conditional mathematical expectation of resulting distribution~\eqref{eq:gp:12} is used as the initial vector-valued function approximation.

We can also take into account the observed errors by assuming that it has the form of an independent identically-distributed Gaussian noise. Then it is easy to show (see~\cite{3569}, section 2.2) that the covariance matrix from equation~\eqref{eq:gp:12} takes the form:
\begin{equation}
\label{eq:gp:13}
\Sigma = \left(\begin{array}{cc}
\Sigma_{\rm d} & \Sigma_{\rm s} \\
\Sigma_{\rm s}^{\rm T} & \Sigma_{\rm d} + \tim{\sigma}^2 I \\
\end{array}\right),
\end{equation}
where $\sigma^2$~-- known variance of Gaussian noise. Equation~\eqref{eq:gp:12} will be rewritten in the form:
\begin{multline}
    \label{eq:gp:14}
    \bm{\upnu_1} | \bm{\upnu_2} \sim N\left( \bm{\mu_1} + \Sigma_{\rm s} \left(\Sigma_{\rm d} + \tim{\sigma}^2 I\right)^{-1}( \bm{\upnu_2} - \bm{\mu_2}), \right. \\
     \left. \Sigma_{\rm d} - \Sigma_{\rm s} \left(\Sigma_{\rm d} + \tim{\sigma}^2 I\right)^{-1} \Sigma_{\rm s}^{\rm T} \right).
\end{multline}

We implemented such an approach as a Python package\footnote{\url{https://github.com/matwey/gp-multistate-kernel}}.
Our implementation is relies heavily on the {\tt sklearn} package. It follows from equation (\ref{eq:gp:7}) that the mentioned approach reuses conventional Gaussian process kernels as nested kernels for describing independent single-variable generating processes. Thus, we only have to implement the evaluation of equation (\ref{eq:gp:7}) and the derivatives to fit the conventional maximal-likelihood process. Reusing {\tt sklearn} package allows us to have the efficient and proven implementation of maximal-likelihood fitting.

In other words, our implementation of the {\tt MultiStateKernel} class is parameterized by arbitrary nested GP kernels from {\tt sklearn}, which is subsequently used as an argument for {\tt sklearn.GaussianProcessRegressor}.

Unfortunately, {\tt sklearn}~1.2 does not support multidimensional regressor output, so additional input data transformation is required. Each output vector coordinate is considered as separate scalar data row and additional integer-valued feature is added to enumerate the coordinate index for this row.

\subsection{Light curve approximation with vector Gaussian processes}
\label{ap_gp}

We applied the vector GPs to the SLSN sample described in Section~\ref{sample}. For each object, GP kernels have been defined as follows:
\begin{equation}
    \label{approx:1}
    k_i(t, t') = \exp\left(-\frac{(t - t')^2}{2l_i^2}\right) 
    \\
    i = 1, \ldots, \tim{n} - 1,
\end{equation}
\begin{equation}
    k_{\tim{n}}(t, t') =  \begin{cases} 0, & t \neq t' \\
    \sigma_i, & t = t' \end{cases},
\end{equation}
where $\sigma_i$-parameter is a noise level, $l_i$-parameter is a length scale, and $\tim{n}$ --- number of passbands available for a certain object.

As noted in the Section~\ref{gp}, the parameters can be found by the maximum likelihood method, moreover, we can impose conditions on the parameter $l$, based on physical considerations. In our case, parameter $l$ is the characteristic time interval between two observations, during which the correlation between these observations can still be considered as significant. In other words, if $l$ is too big, then the GPs will tend to correlate all points on the light curve with each other, as a result of which the model will be smoothed until, in the limiting case, it becomes a constant. On the other hand, the smaller $l$, the lower correlation between two points in time, and the model will tend to pass through all training points, which can cause anomaly peaks and fluctuations in the curve. Therefore, we have limited the parameter $l$ by the condition: $l \in \left[0.5, 100 \right]$.

After approximating multicolor light curves with vector GPs, we excluded those objects from our SLSN sample that did not exhibit a clearly defined peak brightness. Besides, more careful literature search revealed that two objects in our sample are in fact the tidal disruption events: PS16dtm (also known as SN2016ezh, \citealt{2017ApJ...843..106B}) and ASASSN-15lh (also known as SN2015L, 
\citealt{2016NatAs...1E...2L}). 

Thus, 29 SLSNe remained in our sample. They are listed in Table~\ref{tab}. For each supernova, the table contains the equatorial coordinates ($\alpha$, $\delta$), redshift ($z$), passbands in which photometric data are available in the catalogue, and corresponding references. SLSNe are divided into three groups according to assigned subtype: SLSN-I, SLSN-II, SLSN-R.


\begin{table*}
\caption{Sample of SLSNe extracted from the Open Supernova Catalog.}
\label{tab}
\begin{tabular}{lccclr}
\hline
Name & $\alpha$ & $\delta$ & $z$ & Passbands & References \\
\hline \multicolumn{6}{c}{SLSN-I} \\ \hline
SN2018hti & 03:40:53.75 & $+$11:46:37.3 & 0.06 & {\textit{u, g, r, i, B, V}} & {[1]}\\
SN2015bn & 11:33:41.57 & $+$00:43:32.2 & 0.11 & {\textit{r$'$, U, B, V, R, I}} & {[2, 3, 4]} \\
SN2016eay & 12:02:51.71 & $+$44:15:27.4 & 0.10 & {\textit{r, i, U, B, V, R}} & {[3, 5, 6]}\\
SN2010gx & 11:25:46.71 & $-$08:49:41.4 & 0.23 & {\textit{r, u$'$, g$'$, r$'$, i$'$, z$'$}} & {[7, 8, 9]}\\
PTF09cnd & 16:12:08.94 & $+$51:29:16.1 & 0.26 & {\textit{g, r, i, U, B, R}} & {[3, 7, 9]}\\
SN2011kg & 01:39:45.51 & $+$29:55:27.0 & 0.19 & {\textit{g, r, i, z, B}} & {[7, 9, 10]}\\
SN2011ke & 13:50:57.77 & $+$26:16:42.8 & 0.14 & {\textit{u, g, r, i, z, V}} & {[7, 9, 10]}\\
SN2010md & 16:37:47.00 & $+$06:12:32.3 & 0.10 & {\textit{g, r, i, z, B}} & {[7, 9, 10]}\\
PTF12mxx & 22:30:16.73 & $+$27:58:22.0 & 0.33 & {\textit{u, g, r, i, z, B}} & {[3, 7, 9]}\\
PTF09atu & 16:30:24.55 & $+$23:38:25.0 & 0.50 & {\textit{u, g, r, i, U, R}} & {[3, 7, 9]}\\
iPTF13ajg & 16:39:03.95 & $+$37:01:38.4 & 0.74 & {\textit{g, r, i, z, B, R}} & {[5, 11]}\\
LSQ12dlf & 01:50:29.80 & $-$21:48:45.4 & 0.26 & {\textit{U, B, V, R, I}} & {[12, 13]}\\
PTF12gty & 16:01:15.23 & $+$21:23:17.4 & 0.18 & {\textit{g, r, i}} & {[5, 9]}\\
SN2013hy & 02:42:32.82 & $-$01:21:30.1 & 0.66 & {\textit{g, r, i, z}} & {[5, 14, 15, 16]}\\
SN2013dg & 13:18:41.35 & $-$07:04:43.0 & 0.27 & {\textit{g, r, i, z}} & {[12]}\\
LSQ14mo & 10:22:41.53 & $-$16:55:14.4 & 0.26 & {\textit{u, g, r, i, U}} & {[3, 5, 17]}\\
PS1-10bzj & 03:31:39.83 & $-$27:47:42.2 & 0.65 & {\textit{g, r, i, z}} & {[5, 18, 19]}\\
PTF10aagc & 09:39:56.92 & $+$21:43:17.1 & 0.21 & {\textit{g, r, i, U}} & {[3, 5, 9]}\\
DES14X3taz & 02:28:04.46 & $-$04:05:12.7 & 0.61 & {\textit{g, r, i, z}} & {[5, 20]}\\
SNLS-07D2bv & 10:00:06.63 & $+$02:38:35.8 & 1.50 & {\textit{g, r, i, z}} & {[5, 22]}\\
SN2012il & 09:46:12.91 & $+$19:50:28.7 & 0.18 & {\textit{g, r, i, z }} & {[5, 10, 14]}\\
PTF10uhf & 16:52:46.70 & $+$47:36:21.8 & 0.29 & {\textit{g, r, i}} & {[7, 9]}\\
\hline \multicolumn{6}{c}{SLSN-II} \\ \hline
SN2006gy & 03:17:27.06 & $+$41:24:19.5 & 0.02 & {\textit{r$'$, i$'$, U, B, V, R}} & {[3, 5, 24, 25]}\\
PTF12mkp & 08:28:35.09 & $+$65:10:55.6 & 0.15 & {\textit{U, B, V}} & {[3, 5, 7]}\\
\hline \multicolumn{6}{c}{SLSN-R} \\ \hline
LSQ14bdq & 10:01:41.60 & $-$12:22:13.4 & 0.35 & {\textit{g, r, i, z}} & {[21]}\\
PTF12dam & 14:24:46.20 & $+$46:13:48.3 & 0.11 & {\textit{g, r, i, B }} & {[3, 7, 26, 27]}\\
SN2007bi & 13:19:20.19 & $+$08:55:44.3 & 0.13 & {\textit{B, V, R, I}} & {[5, 28, 29]}\\
PS1-14bj & 10:02:08.43 & $+$03:39:19.0 & 0.52 & {\textit{g, r, i, z}} & {[5, 19, 30]}\\
iPTF13ehe & 06:53:21.50 & $+$67:07:56.0 & 0.34 & {\textit{g, r, i}} & {[31]}\\
\hline
\multicolumn{6}{l}{[1]-~\citet{2020MNRAS.497..318L}; [2]-~\citet{2016ApJ...826...39N}; [3]-~\citet{2014Ap&SS.354...89B}; [4]-~\citet{2016ApJ...828L..18N};}\\

\multicolumn{6}{l}{[5]-~\citet{2012PASP..124..668Y}; [6]-~\citet{2017ApJ...835L...8N}; [7]-~\citet{2016ApJ...830...13P}; [8]-~\citet{2010ApJ...724L..16P};}\\

\multicolumn{6}{l}{[9]-~\citet{2017arXiv170801623D}; [10]-~\citet{2013ApJ...770..128I}; [11]-~\citet{2014ApJ...797...24V}; [12]-~\citet{2014MNRAS.444.2096N};}\\

\multicolumn{6}{l}{[13]-~\citet{2015A&A...579A..40S}; [14]-~\citet{1989A&AS...81..421B}; [15]-~\citet{2013ATel.5603....1P}; [16]-~\citet{2015MNRAS.449.1215P};}\\

\multicolumn{6}{l}{[17]-~\citet{2014ATel.5839....1L}; [18]-~\citet{2013ApJ...771...97L}; [19]-~\citet{2018ApJ...852...81L}; [20]-~\citet{2016ApJ...818L...8S};}\\

\multicolumn{6}{l}{[21]-~\citet{2015ApJ...807L..18N}; [22]-~\citet{2013ApJ...779...98H}; [23]-~\citet{1989A&AS...81..421B}; [24]-~\citet{2007ApJ...666.1116S};}\\

\multicolumn{6}{l}{[25]-~\citet{2017ApJS..233....6H}; [26]-~\citet{2017ApJ...835...64G}; [27]-~\citet{2013Natur.502..346N}; [28]-~\citet{2009Natur.462..624G};}\\

\multicolumn{6}{l}{[29]-~\citet{2010A&A...512A..70Y}; [30]-~\citet{2016ApJ...831..144L}; [31]-~\citet{2015ApJ...814..108Y}.} \\

\end{tabular}
\end{table*}

\begin{figure*}
\begin{minipage}{0.47\linewidth}
\center{\includegraphics[width=1\linewidth]{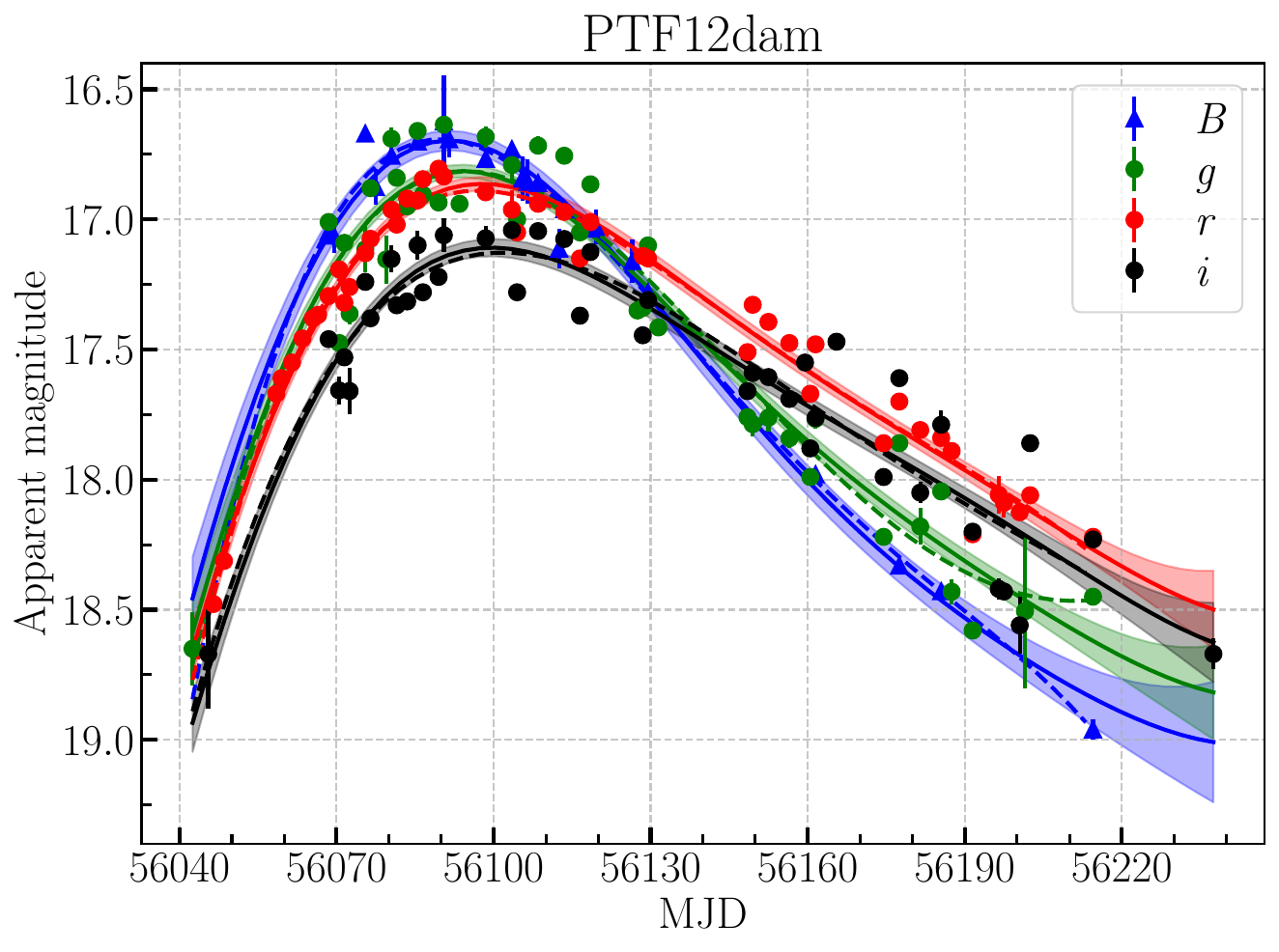}}
\end{minipage}
\hfill
\begin{minipage}{0.47\linewidth}
\center{\includegraphics[width=1\linewidth]{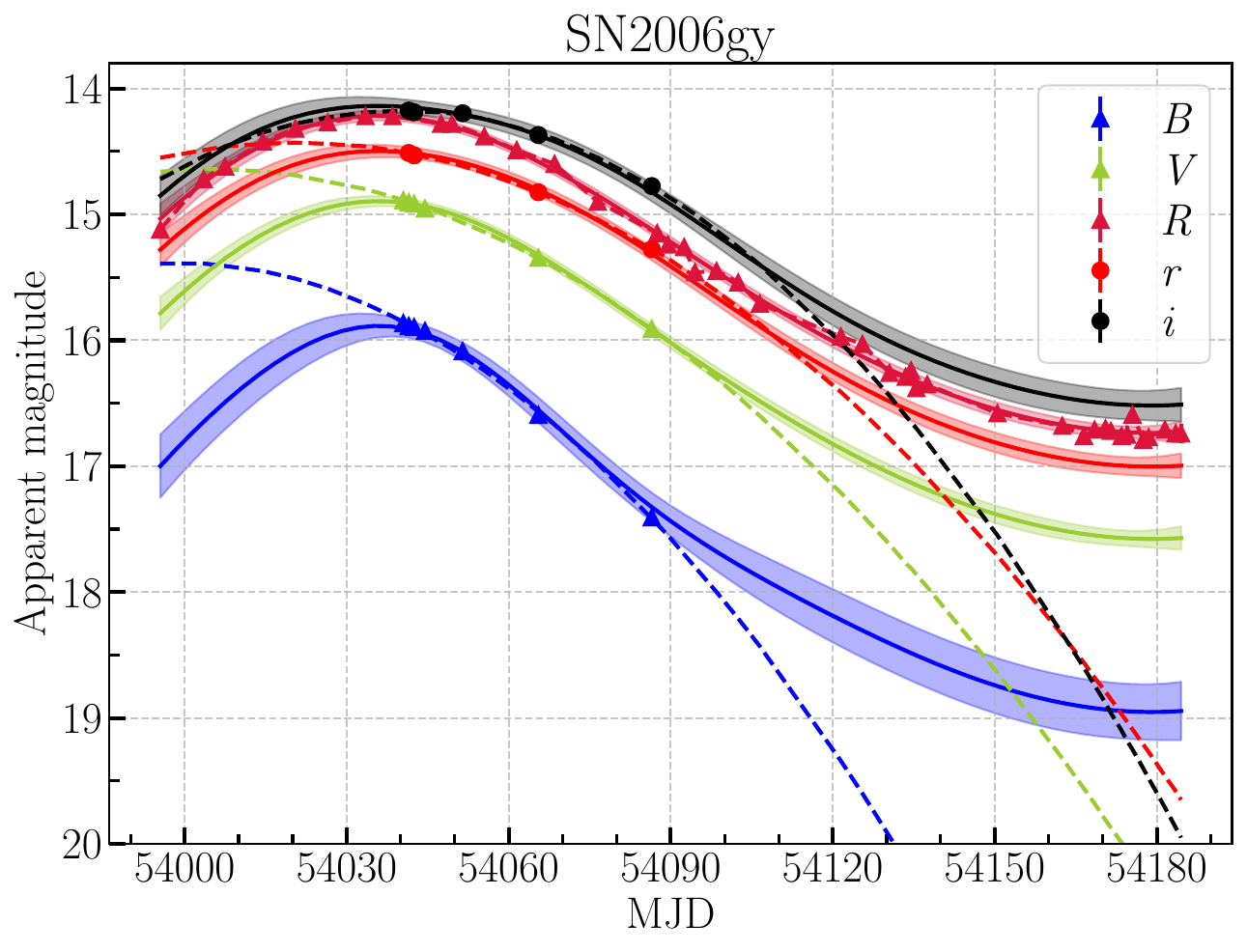}}
\end{minipage}
\caption{\label{approx_multicol1} The result of vector GPs approximation of multicolor light curves of SLSN PTF12dam and SN2006gy. The points represent measurements in the corresponding filters. The solid lines are the estimated mean $\mu$ values. The shaded areas represent $\pm \sigma$ uncertainty. The dashed lines represent the optimal polynomial fit provided by \texttt{SuperBol}~\citep{2018RNAAS...2..230N}.}
\end{figure*}

Fig.~\ref{approx_multicol1} shows the approximation results of PTF12dam and SN2006gy multicolor light curves with vector GPs (solid lines) and with polynomial fit (dashed lines) provided by \texttt{SuperBol} (\citealt{2018RNAAS...2..230N}, Section~\ref{superbol}).
The shaded areas indicate the range of model uncertainty, where the less-likelihood curves are located.
For PTF12dam, the OSC photometric data are fairly uniform in time and, therefore, not difficult to approximate. The advantage of the vector GPs reveals itself when dealing with objects having non-uniform in time photometric data in different passbands (see Section~\ref{simple_approx}). 
For example, for SN2006gy (Fig.~\ref{approx_multicol1}, on the right) none of the LCs (except $R$-band) contains the observations before maximum light. However, the vector GPs successfully approximate these missing parts of the light curves, based on more complete $R$-band observations.
Unfortunately, it is impossible to propose a straightforward physical interpretation of the vector GP model.
For instance, the vector GP model does not simply assume that photometric colors remain constant within the missing part of the curve. 
Instead, the restored part of the curve is evaluated using all available measurements, which are weighted in a sophisticated manner.

The results of multicolor light-curve approximation with the vector GPs for the rest of the sample can be found on \texttt{GitHub}\footnote{\label{slsn-bol}\url{https://github.com/semtim/SLSN-bol}}. 
The obtained approximations are used below to construct the bolometric LCs of SLSNe from our sample.

\section{Construction of bolometric light curves}
\label{bol}

The bolometirc LCs are constructed under assumption that at each moment of time SLSN spectrum can be described by a black body (BB) spectrum with a certain temperature $T$:
\begin{equation}\label{eq:BB}
  B(\nu, T)=\frac{2h\nu^{3}}{c^{2}}\frac{1}{\exp\left(\frac{h\nu}{kT}\right)-1}\,,
\end{equation}
where $\nu$ is frequency, $h$ is Planck's constant, $k$ is Boltzmann's constant, and $c$ is the speed of light. 

Since the objects from our sample are located at different redshift, we have to take into account its effect at the emitted monochromatic flux density. First, photon energies and arrival rates are redshifted, reducing the flux density by a factor of $(1+z)^2$. Counteracting this, the frequency is reduced by a factor of $(1+z)$, leading to a decrease in the flux density by one power of $(1+z)$. Finally, the observed photons at frequency $\nu$ were emitted at frequency $\nu (1+z)$. 

Consequently, building on the BB approximation and accounting for the different magnitude zero-points and the effect of redshift, we derive the following formulae for the apparent magnitudes:

\begin{flalign}
\label{eq:bol:mag}
  & m_{{\rm bb}}^i = -2.5 \log_{10} \left( \frac{\pi R^2 \int {B(\nu (1+z)) \phi_i(\nu) (h\nu)^{-1} d\nu} }{ (1+z) \, D_M^2 \int{3631 {\rm Jy}~ \phi_i(\nu)  (h\nu)^{-1} d\nu}} \right), & \nonumber \\
  & \quad i \in \{u, g, r, i, z\}, & \\
  & m_{{\rm bb}}^i = -2.5 \log_{10} \left( \frac{\pi R^2 \int {B(\nu (1+z)) \phi_i(\nu) d\nu} }{ (1+z) \, D_M^2 \int{f_{\rm Vega}(\nu)~ \phi_i(\nu) d\nu}} \right), & \nonumber \\
  & \quad i \in \{U, B, V, R, I\}, & \nonumber
\end{flalign}
where $R$ denotes the radius of the object, $D_M$ refers to the transverse comoving distance; $f_{\rm Vega}(\nu)$ represents the Vega spectral flux density, $3631 {\rm Jy}$ is the flux zero-point in the AB system. The filter response $\phi_i(\nu)$ is expressed in terms of quantum efficiencies for $ugriz$ passbands, and 1/energy transmission units for $UBVRI$ passbands. This results in a difference in the equations by a $(h\nu)^{-1}$ term.

In order to determine the parameters of the black body ($T$, $R$) at each moment of time, the least squares method is used:
\begin{equation}
    \label{eq:bol:lsm}
    F(T,R) = \sum_{i=1}^{\tim{n}}{\left( m_{\rm gp}^i - m_{\rm bb}^i(T,R)   \right)^2} \rightarrow min,
\end{equation}
where $m_{\rm gp}^i$~--- apparent magnitude in the $i$-th passband of the vector GP approximation (Section~\ref{ap_gp}); $m_{\rm bb}^i(T,R)$~--- $i$-th passband magnitude of the BB model with parameters $T, R$; $\tim{n}$~--- number of passbands for this object.

We numerically minimize the function in equation~\eqref{eq:bol:lsm} to extract the parameters of the black body at each moment for all objects in our sample. This, in turn, allows us to compute the rest-frame bolometric luminosity using the formula:
\begin{equation}\label{eq:bol:7}
  L = 4 \pi R^2 \sigma_\mathrm{SB} T^4,
\end{equation}
where $\sigma_\mathrm{SB}$~is Stefan–Boltzmann's constant.

\begin{figure}
        \center{\includegraphics[width=\columnwidth]{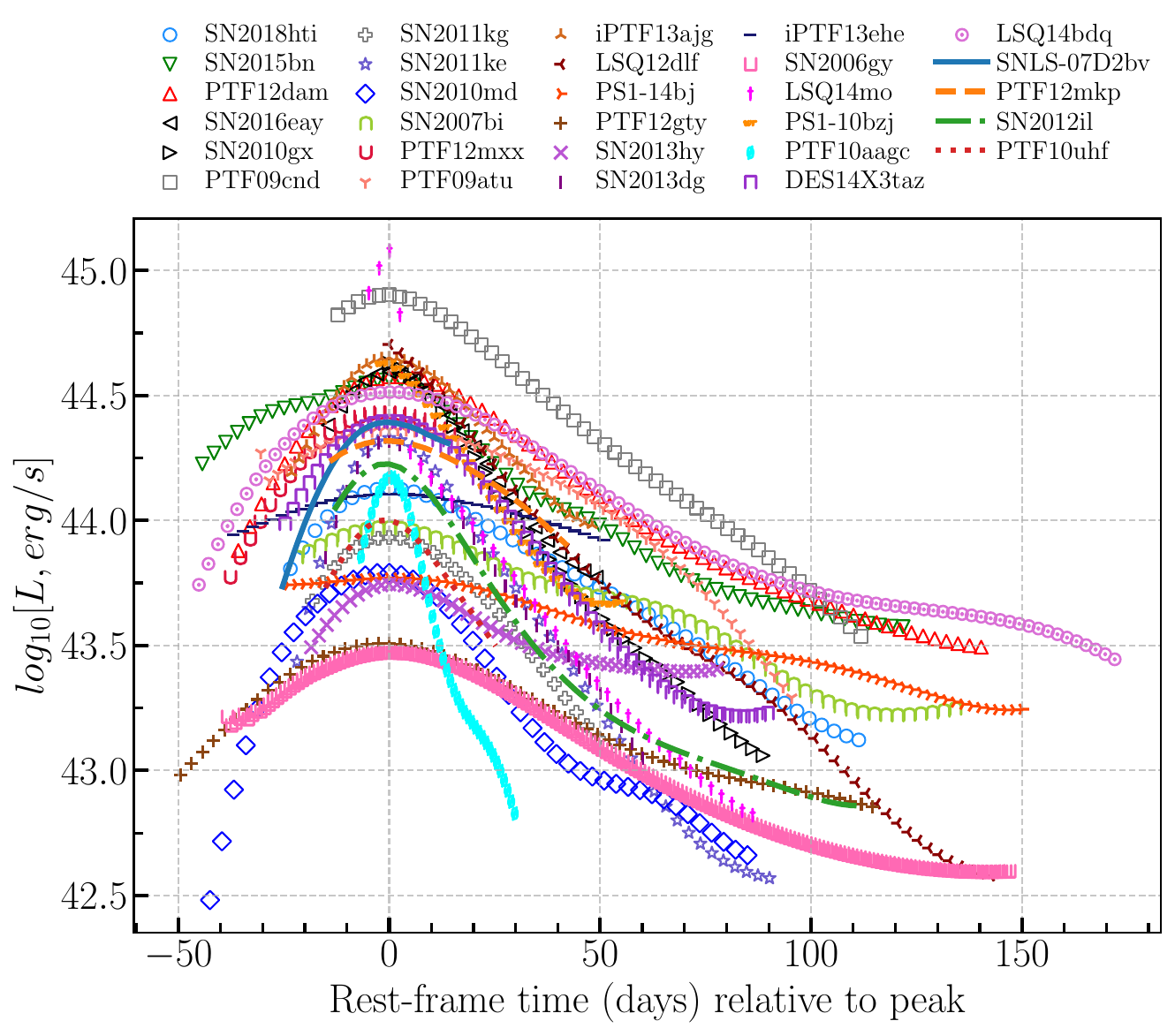}}
        \caption{\label{fig:bol:all} Bolometric light curves in the rest-frame of all SLSNe from our sample.}
\end{figure}

Fig.~\ref{fig:bol:all} shows the bolometric light curves of all SLSN from our sample in the rest frame, with the maximum of light shifted to zero. The tabulated bolometric light curves, as well as graphical representations of the BB radius and temperature variations over time, can be accessed on \texttt{GitHub}\footnoteref{slsn-bol}.

\begin{figure}
        \center{\includegraphics[width=\columnwidth]{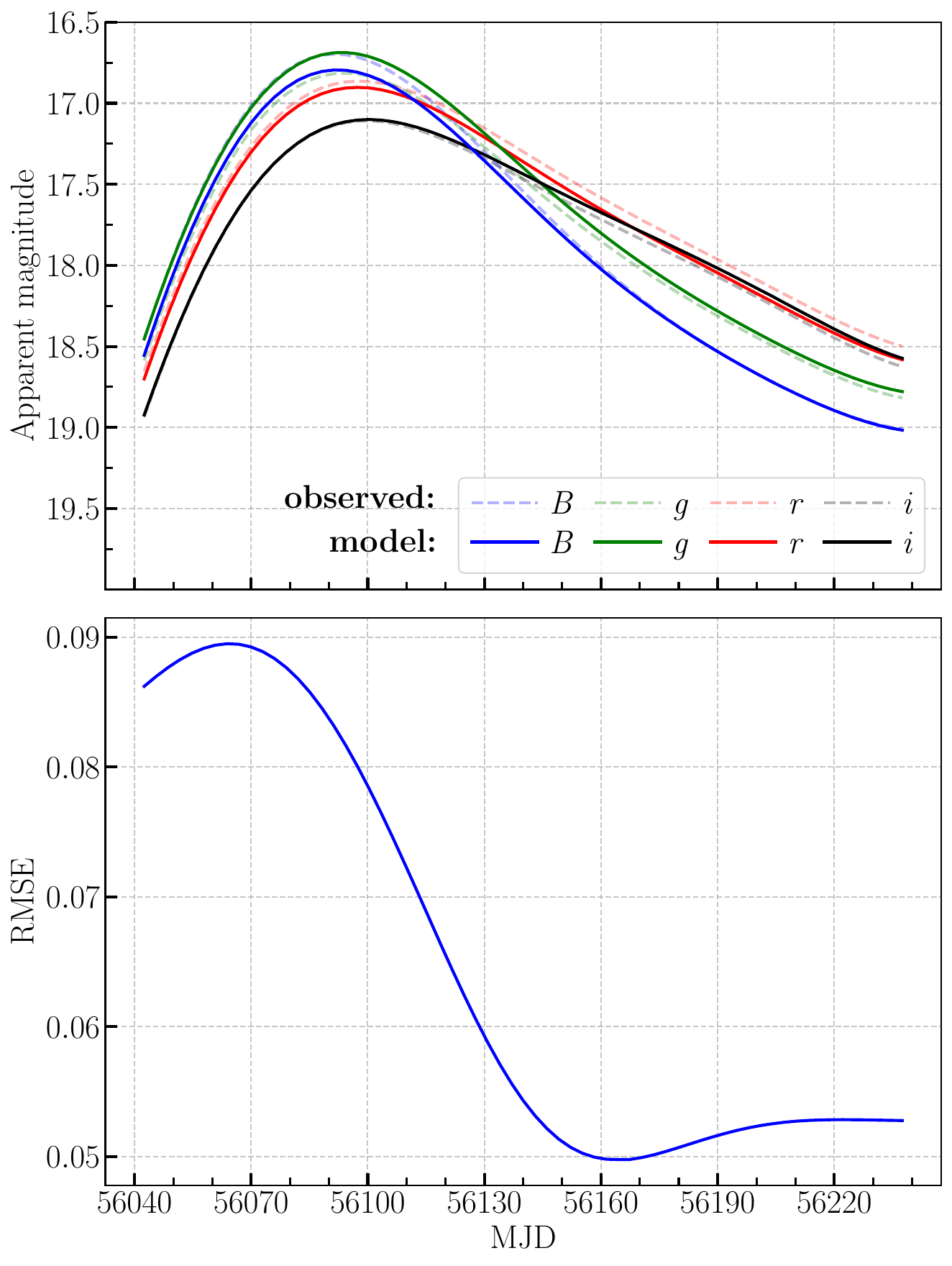}}
\caption{\label{accur} Top: multicolor light curves of PTF12dam in different passbands, approximated by vector GPs (solid lines) and modelled by the BB (dashed lines). Bottom: root mean square error of difference between approximated and modelled magnitudes.}
\end{figure}

Fig.~\ref{accur} demonstrates that the implemented algorithm correctly identifies the BB parameters while minimizing deviations from the approximated light curves, using PTF12dam as an example. At the top of the figure, the approximated light curves are compared with the ones corresponding to the Planck spectrum calculated with equations~\eqref{eq:bol:mag} in the same passbands. It can be noticed that the model light curves have only minor deviations from the approximated curves. At the bottom of the figure, there are the root mean square errors (RMSE) of the predicted apparent magnitudes at each moment of time, corresponding to the identified model parameters. As illustrated, the RMSE does not exceed $0.09$. 

\section{Discussion}
\label{disc}

\subsection{Comparison with traditional Gaussian processes}
\label{simple_approx}

\begin{figure}
        \center{\includegraphics[width=\columnwidth]{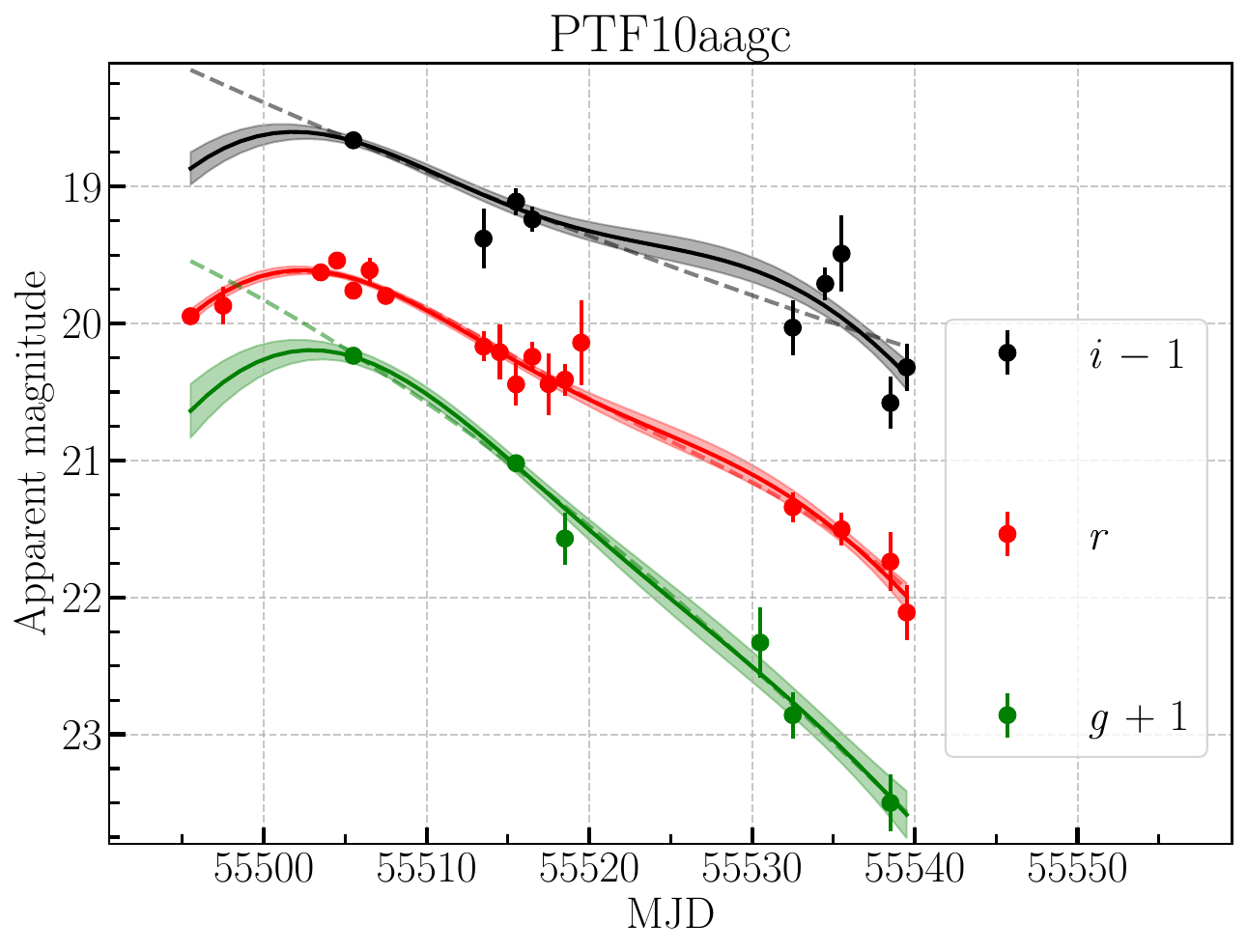}}
\caption{\label{fig:comp} Multicolor light-curve approximation for PTF10aagc by traditional (dashed lines) and vector (solid lines) Gaussian processes. The shaded areas represent $\pm \sigma$ uncertainty of vector GPs.}
\end{figure}

As previously discussed (see Section~\ref{overview}), light curves can be approximated using traditional GPs for each filter separately. However, this method can potentially produce artifacts in the approximation of time-inhomogeneous light curves. For instance, Fig.~\ref{fig:comp} presents the results of LC approximation by traditional and vector GPs for the object PTF10aagc. It can be seen that in the $r$ passband with a better data coverage, traditional GPs provide a satisfactory approximation. However, in the $g$ and $i$ passbands, where data points are few and unevenly distributed, the result of traditional GPs is unsatisfactory, rendering such curves unsuitable for processing by this method. On the other hand, vector GPs demonstrate their efficacy in approximating such ``incomplete'' data. The most significant outcome is the ability to restore a luminosity peak in all fitted light curves. This is important because many approximation methods tend to ``smooth out'' or flatten the light curves before a peak if there are no observations leading up to that peak.

\subsection{Comparison with {\tt SuperBol}}
\label{superbol}

\begin{figure*}
\begin{minipage}{0.47\linewidth}
\center{\includegraphics[width=1\linewidth]{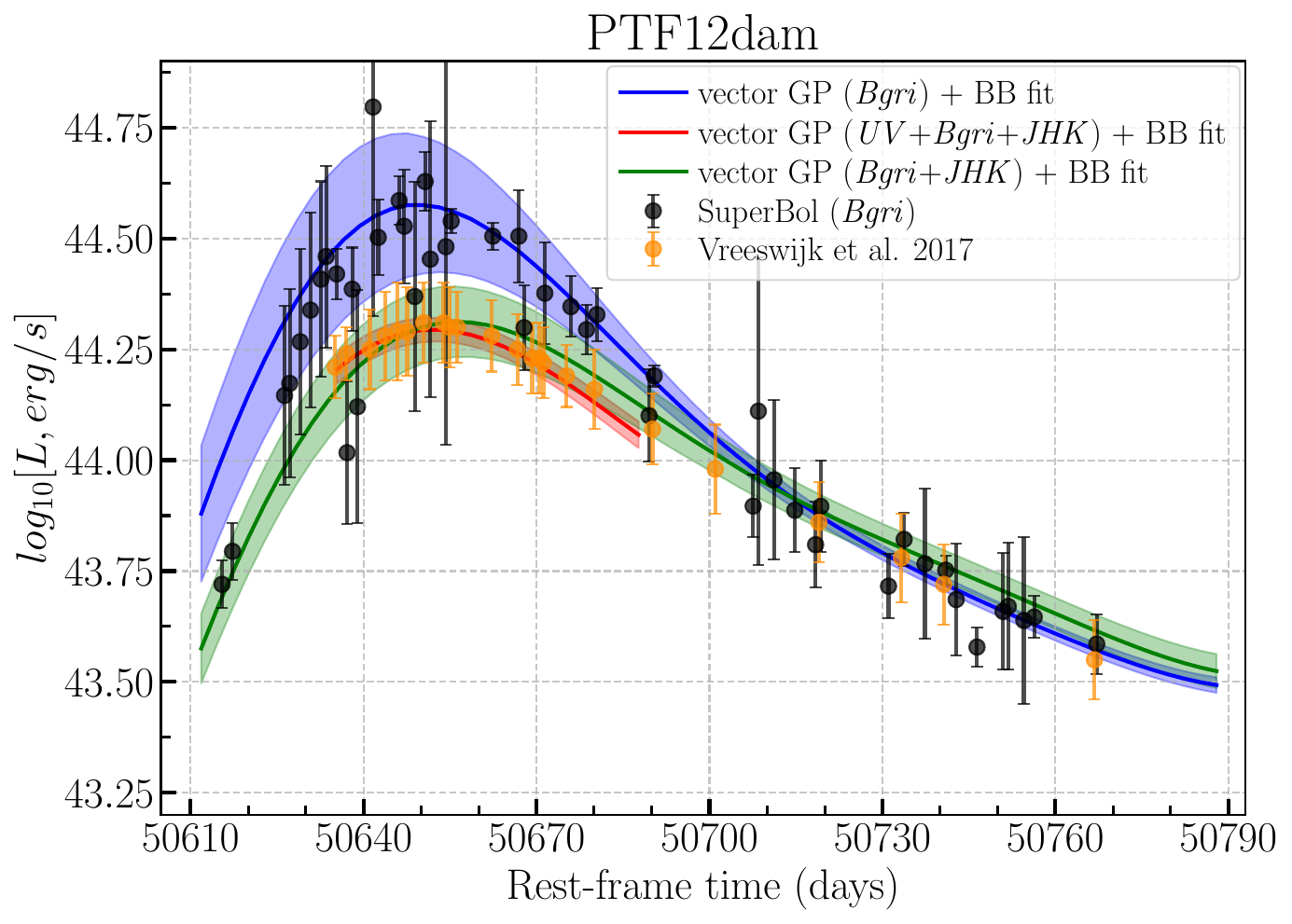}}
\end{minipage}
\hfill
\begin{minipage}{0.47\linewidth}
\center{\includegraphics[width=1\linewidth]{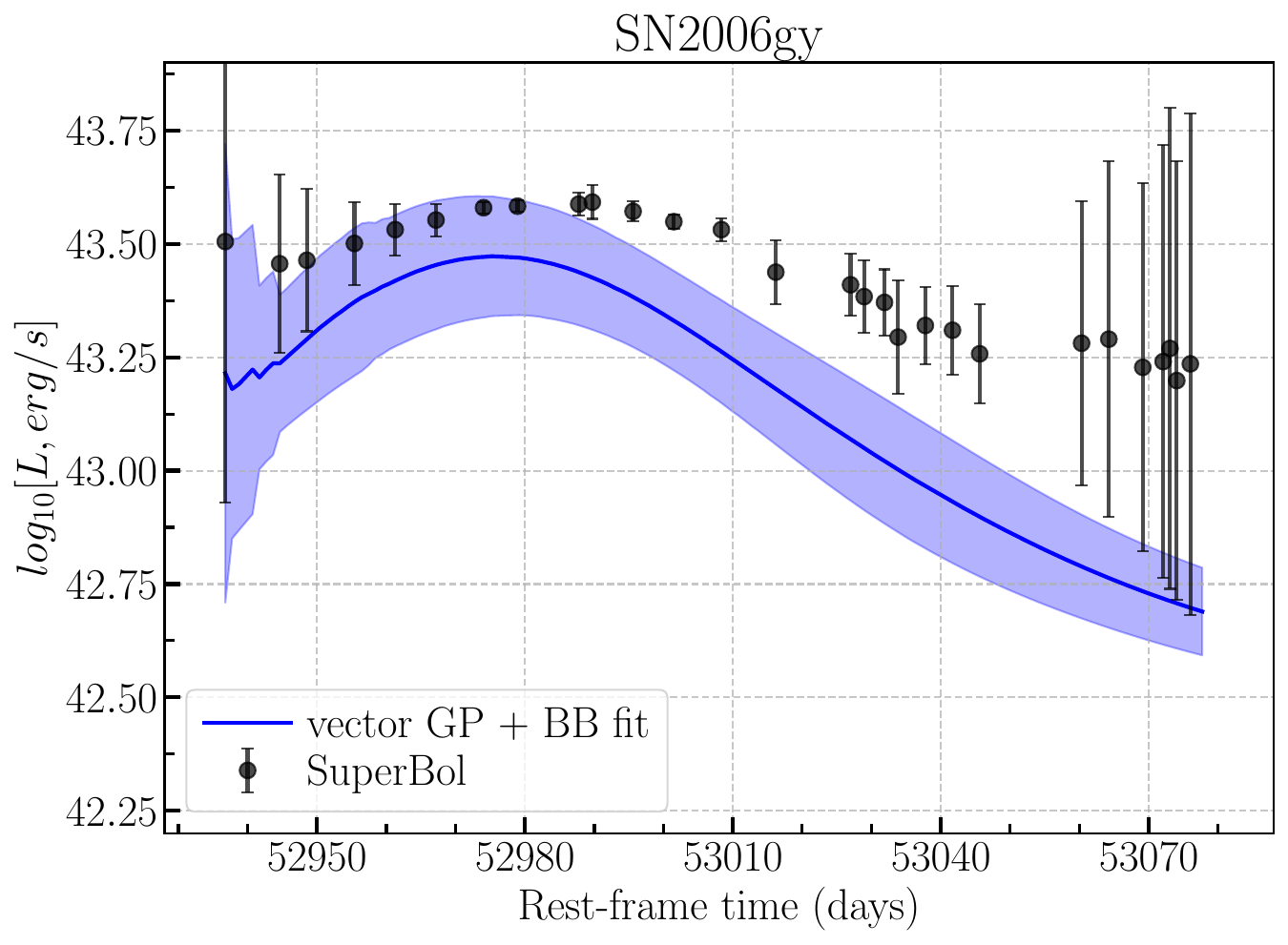}}
\end{minipage}
\caption{\label{fig:concl:1} Bolometric light curves of PTF12dam and SN2006gy. The solid lines with shaded areas represent the black body fit, based on the vector GP approximation, along with its $\pm \sigma$ uncertainty. Black circles indicate the light curve constructed with {\tt SuperBol} using polynomial fits. The light curve obtained by \citet{Vreeswijk_2017} for PTF12dam is marked with orange circles; the colours of the BB fit signify different subsets of photometric passbands used for light-curve approximation.}
\end{figure*}

We compared our approach with {\tt SuperBol} --- a Python code to construct bolometric light curves from time-series multicolour photometry~\citep{2018RNAAS...2..230N}. {\tt SuperBol} also assumes a BB spectrum to account for the flux not covered by the observations, but it uses a polynomial fit to interpolate and extrapolate light curves to a common set of reference times, and each light curve is treated independently.  Fig.~\ref{fig:concl:1} shows the bolometric light curves of superluminous supernova PTF12dam and SN2006gy constructed by different methods using the same photometric data. The blue curve represents our methodology using vector Gaussian processes, while the black points denote the fit conducted with the {\tt SuperBol} code. For well-sampled PTF12dam LCs, both approaches align favorably.  On the other hand, for SN2006gy, which has poor photometric coverage, the {\tt SuperBol} bolometric LC tends to lie above ours, especially in the later stages. This can be attributed to the inferior performance of polynomial fits as compared to vector Gaussian processes (see Fig~\ref{approx_multicol1}).

{\tt SuperBol} has been employed for systematic studies of SLSN light-curve timescales and shape properties (e.g.,~\citealt{2015MNRAS.452.3869N}). \cite{Chatzopoulos_2019} also used {\tt SuperBol} to construct bolometric light curves of SLSNe from the OSC. However, their analysis was focused on the data with robust LC coverage. As mentioned earlier, polynomial functions used in {\tt SuperBol}, may not be as efficient during extrapolation compared to the vector Gaussian processes. Leveraging this advantage, we have managed to expand the sample size of SLSNe in our study ($N_{\rm this~work}$ = 29 vs. $N_{\rm Chatzopoulos~et~al.}$ = 25).

\subsection{Remaining questions and further improvements}

While constructing the bolometric light curves, we simplified our approach by assuming that SLSN spectrum is described by a black body model. This is a reasonable approximation for the optical to near-infrared (NIR) wavelengths (e.g., \citealt{2017ApJ...835L...8N}), though the NIR range can also demonstrate significant deviations from the BB approximation, as, for instance, is shown by \citet{Vreeswijk_2017} for PTF12dam. Nonetheless, the ultraviolet (UV) part of the spectrum experiences significant absorption by heavy elements (e.g.,~\citealt{2017ApJ...840...57Y,10.1093/mnras/stz1321}). Alternatively, there can be a flux excess instead of blanketing in the case of superluminous interacting supernovae (e.g., \citealt{2015MNRAS.449.4304D}). Therefore, using a modified black body model that takes into account the UV part could improve the further modelling. To show the difference, Fig.~\ref{fig:concl:1} includes the bolometric light curve of PTF12dam obtained by \citet{Vreeswijk_2017}, who employed a more sophisticated model with an additional parameter to scale the strength of absorption features in the ultraviolet range.

To evaluate the contribution of UV and NIR parts to the bolometric light curve of PTF12dam, we supplemented the OSC data with UV observations from \textit{Swift} \citep{Nicholl_2013} and NIR photometry in $JHK$-bands as provided by \citet{Vreeswijk_2017}. The duration of the bolometric LC was constrained by the availability of UV photometry, specifically $-30$ and $+30$ days relative to the $r$-band maximum in the observer frame. As illustrated in Fig.~\ref{fig:concl:1}, in this case, our method produces results consistent with those of \citet{Vreeswijk_2017}. We also extended the light curve to later times using $Bgri$ photometry from OSC and $JHK$ photometry. As can be seen from Fig.~\ref{fig:concl:1}, at later times ($\sim$40 days after the peak), all light curves behave similarly.

\section{Conclusions}
\label{concl}

In this study, we focus on reconstructing the bolometric light curves of supernovae, particularly superluminous supernovae. To achieve this, we assembled a sample of 29 SLSNe from the Open Supernova Catalog. Their multicolor light curves were then approximated using vector Gaussian processes, a novel methodology introduced by this study. With the application of the black-body assumption, we reconstructed the bolometric light curves of these SLSNe.

The application of vector GPs distinguishes itself from traditional methods such as polynomial interpolation and standard GPs by taking into account the correlation between light curves in different passbands. This consideration particularly benefits the analysis of data that are sparse or irregularly sampled, thereby potentially enhancing the accuracy of bolometric light curve construction.

Our proposed approach utilizing vector GPs extends beyond supernovae and has the potential to inform studies of various astrophysical transients, including but not limited to exotic phenomena like kilonovae and tidal disruption events. In anticipation of the abundant light curve data from forthcoming projects like the Vera Rubin Observatory Legacy Survey of Space and Time\footnote{\url{https://www.lsst.org/}} or the Nancy Grace Roman Space Telescope\footnote{\url{https://www.stsci.edu/roman}}, the development of advanced tools for multi-output regression analysis becomes crucial. This study serves as a stepping stone in that direction, proposing the use of vector GPs as a viable solution for handling future astronomical datasets.

\section*{Acknowledgements}
We acknowledge Konstantin Malanchev and Patrick Aleo for reading the manuscript and providing insightful comments.

The reported study was funded by RFBR and CNRS according to the research project №21-52-15024. The authors acknowledge the support by the Interdisciplinary Scientific and Educational School of Moscow University “Fundamental and Applied Space Research”. We used the equipment funded by the Lomonosov Moscow State University Program of Development. T.S. is supported by Nonprofit Foundation for the Development of Science and Education ``Intellect''.

\section*{Data Availability}
The vector Gaussian Processes are implemented in the Python package {\tt gp-multistate-kernel}\footnoteref{vgp}. The code written to address the described problem, along with figures of the approximated light curves and the bolometric light curves, are available on \texttt{GitHub}\footnoteref{slsn-bol}. All CSV tables containing the SLSN observations were taken from the Open Supernova Catalog\footnoteref{osc}.



\bibliographystyle{mnras}
\bibliography{ref} 




\appendix

\section{Auxiliary results from the probability theory}

\label{ap:1}
Let $p(\bm{x})$ be probability density function for $k$-dimensional normally distributed vector $\bm{x}$ with mean ${\bm{\mu}}$ and covariance matrix $\Sigma$.
Let
\begin{equation}
    p_{1}(\bm{x_1}) \equiv \int p(\bm{x}) dx_{\tim j}
 \end{equation}
be marginal probability density function for $(k-1)$-dimensional random vector $\bm{x_1}$ obtained by excluding $\tim{j}$-th component from $\bm{x}$.
Then, $p_{1}$ is a probability density function for normal distribution with mean ${\bm{\mu_1}}$ and covariance matrix $\Sigma_{11}$, where both of them are obtained by excluding $\tim{j}$-th row and $\tim{j}$-th column from the initial parameters.
Note that without loss of generality, we may consider $\tim{j}=k$, or reenumerate indices otherwise.
Then $\bm{x} = \left(\begin{aligned}
\bm{x_1} \\
x_k \\
\end{aligned}\right)$,
$\Sigma = \left(\begin{array}{cc}
\Sigma_{11} & \Sigma_{1k}^{T} \\
\Sigma_{1k} & \Sigma_{kk} \\
\end{array}\right)$
and $\Sigma^{-1} = \left(\begin{array}{cc}
\Sigma_{11}^{-1} + \Sigma_{11}^{-1} \Sigma_{1k} H^{-1} \Sigma_{1k}^{T} \Sigma_{11}^{-1} &
-\Sigma_{11}^{-1} \Sigma_{1k} H^{-1} \\
-H^{-1} \Sigma_{1k}^{T} \Sigma_{11}^{-1} &
H^{-1}\\
\end{array}\right)$ by Frobenius block matrix inversion formula, where $H \equiv \Sigma_{kk} - \Sigma_{1k}^{T}\Sigma_{11}^{-1}\Sigma_{1k}$.
It is easy to see the following equation
\begin{multline}
\left(\bm{x}-\bm{\mu}, \Sigma^{-1} \left(\bm{x}-\bm{\mu}\right)\right) = \\
\left(x_k - \mu_k - \Sigma_{1k}^{T}\Sigma_{11}^{-1} \left(\bm{x_1}-\bm{\mu_1}\right), H^{-1}\left( x_k - \mu_k - \Sigma_{1k}^{T}\Sigma_{11}^{-1} \left(\bm{x_1}-\bm{\mu_1}\right) \right)\right) \\
+ \left(\bm{x_1}-\bm{\mu_1}, \Sigma_{11}^{-1} \left(\bm{x_1}-\bm{\mu_1}\right)\right).
\end{multline}
Then,
\begin{equation}
p_{1}(\bm{x_1}) = C \exp\left(-\frac{1}{2}\left(\bm{x_1}-\bm{\mu_1}, \Sigma_{11}^{-1} \left(\bm{x_1}-\bm{\mu_1}\right)\right)\right),
\end{equation}
where constant $C$ can be calculated from probability density function norming constraint.
q.e.d.

\onecolumn
\section{Covariance matrix with independent components}
\label{ap:2}
The special case of covariance matrix for vector $\nu$, with $k$ independent components, in two points of time $t_1$ and $t_2$:
\begin{equation*}
\Sigma = \left(\begin{array}{cccccccc}
\sigma_{1}^2              & 0                         & \dotsm & 0 &
\sigma_{1}^2 K_1(t_1,t_2) & 0                         & \dotsm & 0 \\

0                         & \sigma_{2}^2              & \dotsm & 0 &
0                         & \sigma_{2}^2 K_2(t_1,t_2) & \dotsm & 0 \\

\vdots                    & \vdots                    & \ddots & \vdots &
\vdots                    & \vdots                    & \ddots & \vdots \\

0                         & 0                         & \dotsm & \sigma_{k}^2 &
0                         & 0                         & \dotsm & \sigma_{k}^2 K_k(t_1,t_2) \\

\sigma_{1}^2 K_1(t_1,t_2) & 0                         & \dotsm & 0 &
\sigma_{1}^2              & 0                         & \dotsm & 0 \\

0                         & \sigma_{2}^2 K_2(t_1,t_2) & \dotsm & 0 &
0                         & \sigma_{2}^2              & \dotsm & 0 \\

\vdots                    & \vdots                    & \ddots & \vdots &
\vdots                    & \vdots                    & \ddots & \vdots \\

0                         & 0                         & \dotsm & \sigma_{k}^2 K_k(t_1,t_2) &
0                         & 0                         & \dotsm & \sigma_{k}^2 \\
\end{array}\right),
\end{equation*}
where $K_i(t_1,t_2)$ is called the kernel in the Gaussian process approximation framework.
One may see that this case can be covered by performing $k$ independent approximation procedures.

\bsp	
\label{lastpage}
\end{document}